\documentclass[prl,aps,twocolumn,amssymb,showpacs,superscriptaddress]{revtex4}
\usepackage{graphicx}
\usepackage{color}
\usepackage{times}

\begin{document}

\title{Bragg polaritons: Strong coupling and amplification in an unfolded microcavity}

\author{A.~Askitopoulos$^{1}$, L.~Mouchliadis$^{1,2}$, I.~Iorsh$^{3}$, G.~Christmann$^{4}$, J.~J.~Baumberg$^{4}$, M.~A.~Kaliteevski$^{3}$, Z.~Hatzopoulos$^{1,2}$, and P.~G.~Savvidis$^{2,5}$}
\email[corresponding author: ]{psav@materials.uoc.gr}

\affiliation{$^{1}$Department of Physics, University of Crete,
71003 Heraklion, Crete, Greece\\ $^{2}$IESL-FORTH, P.O. Box 1527,
71110 Heraklion, Crete, Greece\\ $^{3}$Department of Physics,
Durham University, Durham DH1 3BH, United
Kingdom\\ $^{4}$Cavendish
Laboratory, University of Cambridge, Cambridge CB3 0HE, United
Kingdom\\ $^{5}$Department of Materials Science \& Technology,
University of Crete, Greece}

\begin{abstract}
Periodic incorporation of quantum wells inside a one--dimensional
Bragg structure is shown to enhance coherent coupling of excitons
to the electromagnetic Bloch waves. We demonstrate strong coupling
of quantum well excitons to photonic crystal Bragg modes at the
edge of the photonic bandgap, which gives rise to mixed Bragg
polariton eigenstates. The resulting Bragg polariton branches are
in good agreement with the theory and allow demonstration of Bragg
polariton parametric amplification.
\end{abstract}

 \maketitle

Semiconductor microcavities in the strong coupling regime
\cite{weisbuch} offer an ideal testbed for harnessing
light--matter interactions at nanometer scale. In addition, they
appear to be highly suitable systems for realization of a new
generation of optoelectronic devices \cite{tsintzos1,liew} of
unprecedented efficiency \cite{christmann} operating on new
physical principles such as, e.g., polariton lasing
\cite{christopoulos} which requires no population inversion
\cite{imamoglu,deng}. However, the strength of light--matter
coupling in GaAs--based microcavities, normally quantified in
units of Rabi energy, is limited by the number of quantum wells
(QWs) that can be integrated at the antinodes of the electric
field inside the microcavity \cite{savona}. Conventional
approaches to enhance Rabi splitting required for high temperature
operation of these devices include increasing cavity thickness to
incorporate larger number of QWs at the expense of growing mode
volume \cite{pelekanos}. Additionally, to compensate for the
undesired penetration of the electric field into the mirrors,
insertion of extra QWs in the first periods of the Bragg reflector
where electric field retains its strength, has also been
implemented \cite{bloch}.

In recent years the \emph{Bragg polariton} concept has attracted
increasing attention, as a new tool for tailoring light--matter
coupling. Several recent reports explore the possibility of
creating Bragg polaritons by incorporation of excitonic resonances
periodically throughout a photonic crystal. Both confined excitons
inside a QW \cite{faure} as well as large oscillator strength bulk
excitons in materials such as ZnO have been
considered\cite{biancalana}. However, the relatively small
refractive index contrasts of the considered structures impose
limitations on the width of the PBG, hindering clear separation of
Bragg polariton branches \cite{goldberg}. Similarly, exciton-Bragg
mode hybridization has also been reported in a CdTe microcavity
with exciton coupling occurring to both cavity and PBG leaky modes
owing to the very large exciton binding energies of CdTe
systems\cite{richard}.

In this Letter we use a periodic Bragg structure referred to as
{\it unfolded microcavity} with well defined PBG and sharp Bragg
resonances, to experimentally demonstrate strong coherent coupling
between Bragg photons and InGaAs QW excitons. The new eigenmodes
of the system exhibit characteristic anticrossing behavior and are
an admixture of excitons and Bragg photons. The corresponding
Bragg polariton branches are shown to be in excellent agreement
with the theoretical models.

Our structure is an optimized unfolded microcavity in which QWs
are incorporated into a Bragg mirror stack. The sample shown in
Figure 1(c) consists of a stack of quarter wavelength thick
alternating high refractive index ($n_{{\rm GaAs}}=3.5$) GaAs
layer and of a smaller effective refractive index ($n_{{\rm eff}}
\sim 3.2$) pseudo-layer AlAs/GaAs/InGaAs/GaAs/AlAs whose effective
optical thickness is $\lambda/4$. The 10nm wide
In$_{0.1}$Ga$_{0.9}$As QWs are placed symmetrically inside the
pseudo-layer at the antinodes of the electric field as shown in
Figure 1(a). The sample is mounted inside a He cryostat and the
temperature is maintained at 32K.

In Figure 1(b) reflectivity spectra of the pseudo--Bragg stack
without excitonic absorption calculated using standard transfer
matrix technique are presented showing 81meV (50nm) wide PBG
centered at 1.418eV (875nm). The maximum reflectivity at the
center of the PBG for a 30--period stack sample exceeds $99\%$.
Clear indication of the existence of the Bragg polariton mode is
seen in Figure 1(b) (solid red line) when ``switching on'' the
excitonic absorption in the transfer matrix model, which results
in a sharp mode inside the otherwise forbidden PBG. We
experimentally probe Bragg polariton branches using white light
reflectivity measurements. Tuning of the Bragg mode, located at
the high energy side of the PBG through the excitonic resonance is
achieved by scanning the reflectivity spot across the sample at
normal incidence and exploiting the built--in wedge, intentionally
introduced during the sample growth.
\begin{figure}[t]
\includegraphics[width=8.5cm]{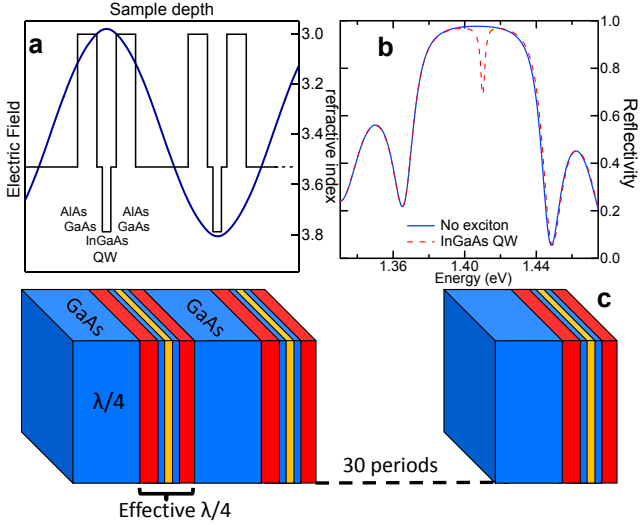}
\caption{(color online) (a) Electric field along the structure
showing InGaAs QW positioned at the electric field maxima. (b)
Reflectivity spectrum in the presence (dashed line) and absence
(solid line) of an exciton resonance showing Bragg polariton mode
(c) Periodic sample structure.}
\end{figure}

In Figure 2(a) a series of reflectivity spectra taken at different
positions across the sample in the order of increasing layer
thickness is presented. Increasing layer thickness results in the
shifting of the whole PBG to lower energies with the corresponding
tuning of the Bragg resonance. It is convenient to define as
detuning $\Delta$, the difference in energy between the edge of
the PBG and the exciton resonance. As seen in Figure 2(a), for
positive detuning conditions a sharp exciton--like resonance is
observed within the PBG while the Bragg mode appears on the high
energy side of the PBG. With decreasing detuning the Bragg mode is
continuously tuned across the excitonic resonance exhibiting a
characteristic anticrossing behavior at zero detuning until it
reappears on the other side of the excitonic resonance. Such
anticrossing behavior with a Rabi splitting of 9.3 meV is a direct
manifestation of the strong coupling regime which gives rise to
the new eigenmodes of the system, namely Bragg polaritons, part
light part matter quasiparticles. Persistence of strong coupling
regime in this sample up to a temperature of 100K has been
experimentally verified.
\begin{figure}
\includegraphics[width=8cm]{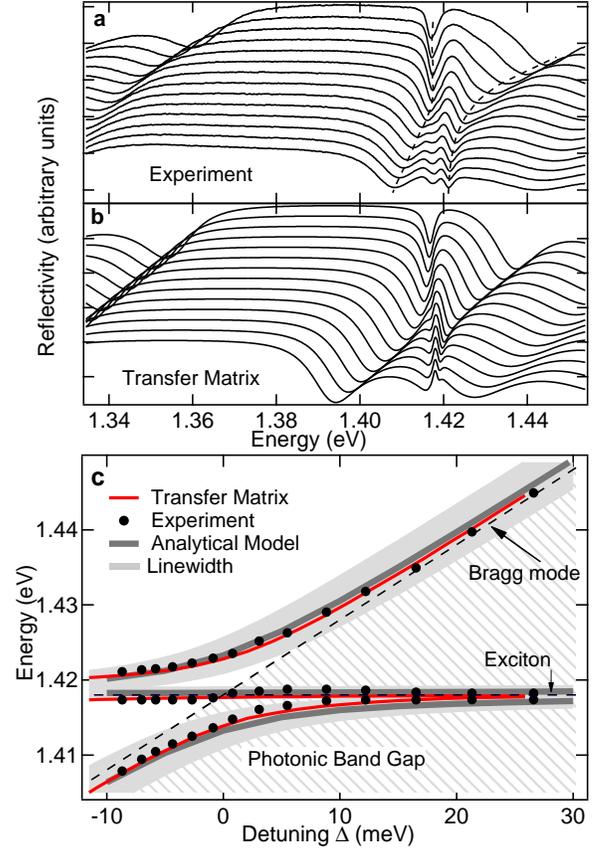}
\caption{(Color online) (a) Experimental reflectivity spectra vs
sample position showing clear anticrossing behavior with Rabi
splitting of 9.3meV. (b) Modeled reflectivity spectra (c)
Extracted Bragg polariton branches vs detuning. Experimental
reflectivity dip positions (black circles), transfer matrix (solid
line) and coupled harmonic oscillator model (thick gray line).
Dashed lines indicate the position of uncoupled exciton and Bragg
modes. Calculated Bragg polariton linewidths are plotted in gray
assuming exciton, photon decay rates 1.5meV, 8meV respectively.}
\end{figure}

Notably, in addition to the two anticrossing branches, a third
dispersionless feature is also observed between them. In order to
shed light on the origin of this dispersionless branch, we perform
transfer matrix calculations of the normal incidence reflectivity
spectra with increasing layer thickness. In Figure 2(b), the
corresponding modeled reflectivity spectra are presented, showing
excellent agreement with the experimental data and reproducing the
anticrossing behavior and the existence of the middle branch. The
good agreement is clearly confirmed when extracting both
experimental (black circles) and calculated (solid line)
reflectivity dip positions for different exciton--Bragg mode
detunings, as shown in Figure 2(c). The position and
non--dispersive nature of the middle branch suggests the presence
of uncoupled excitonic states in the system. In order to confirm
this hypothesis we developed a coupled harmonic oscillator model
to fit our Bragg polariton branches \cite{vardeny, panzarini,
ivchenko}.
\begin{figure}
\includegraphics[width=8cm]{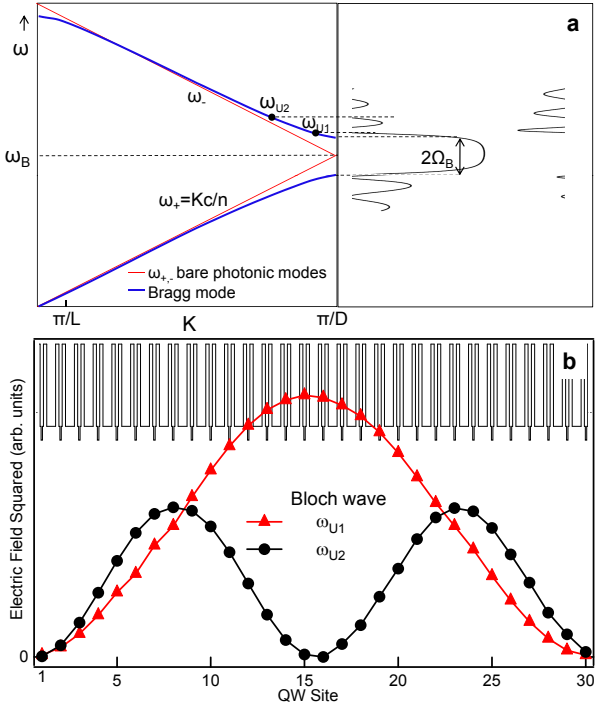}
\caption{(Color online) (a) 1-D photonic crystal band diagram with
respective reflectivity spectra of the considered finite Bragg
structure of length $L$. The first and second Bragg modes are
indicated with $\tilde{\omega}_{U1}$ and $\tilde{\omega}_{U2}$
respectively (b) Electric field squared calculated at the QW sites
for the first (triangles) and second (circles) Bragg modes.}
\end{figure}

We note that our 30--period Bragg structure can be approximated by
an infinite periodic medium characterized by an effective
refractive index $n_{{\rm eff}}$, in which both photonic and
excitonic states are represented using Bloch waves. We denote the
bare uncoupled photonic modes propagating in positive and negative
directions with $|C_{+}\rangle$ and $|C_{-}\rangle$, respectively,
and the corresponding excitonic modes with $|X_{+}\rangle$ and
$|X_{-}\rangle$. Thus, the Hamiltonian of the system can be
written in the form:
\begin{equation}
\mathcal{H} =
 \left( \begin{array}{cccc}
\omega_{+}& \Omega_{{\rm B}} & \Omega_{{\rm R}}& 0\\
\Omega_{{\rm B}} & \omega_{-}& 0 & \Omega_{{\rm R}}\\
\Omega_{{\rm R}}& 0& \omega_{{\rm x}} & 0\\
0 & \Omega_{{\rm R}} & 0 & \omega_{{\rm x}}
\end{array} \right) \label{hamiltonian1}
\end{equation}
where, $\omega_{{\rm x}}$ is the resonance frequency of two
degenerate exciton modes $|X_{+}\rangle$ and $|X_{-}\rangle$,
$\omega_{{\rm B}}$ is the Bragg frequency corresponding to the
center of PBG, $\Omega_{{\rm B}}$ is the half--width of PBG and
$\Omega_{{\rm R}} $ is the vacuum Rabi splitting measuring the
strength of the exciton--photon interaction in the structure. It
is worth noticing that the two exciton modes are degenerate
because excitons in different QWs are well separated and do not
interact directly with each other. As shown in Figure 3, the
energies of the uncoupled photonic modes $|C_{+}\rangle$ and
$|C_{-}\rangle$ can be expressed in terms of the Bloch wavevector
$K$ as $\omega_{+}= K c/n_{{\rm eff}}$ and $\omega_{-}= 2\pi
c/(n_{{\rm eff}}D)-Kc/n_{{\rm eff}}$, where $D$ is the period of
the structure. The value of the Bragg frequency is given by
$\omega_{{\rm B}}= \pi c/ (n_{{\rm eff}} D)$, the parameter which
is varied during the experiment by varying the period of the
structure. The Hamiltonian (\ref{hamiltonian1}) can be simplified
if instead of the bare photonic and excitonic basis states, we use
the basis of upper and lower photonic Bloch modes $|B_{U}\rangle$
and $|B_{L}\rangle$ -- obtained by direct diagonalization of the
photonic part -- and symmetric and antisymmetric combination of
excitonic Bloch states $|X_{S}\rangle$ and $|X_{A}\rangle$.
Expressed in the new basis, the Hamiltonian (\ref{hamiltonian1})
has a simpler form
\begin{equation}
\mathcal{H} =
 \left( \begin{array}{cccc}
\omega_{U}& 0 & 0& \sqrt{2}\Omega_{{\rm R}}\\ 0& \omega_{L} &
\sqrt{2}\Omega_{{\rm R}} & 0\\ 0& \sqrt{2}\Omega_{{\rm R}}&
\omega_{{\rm x}} & 0\\ \sqrt{2}\Omega_{{\rm R}} & 0 & 0 &
\omega_{{\rm x}}
\end{array} \right)\,, \label{hamiltonian2}
\end{equation}
where $\omega_{U}$  and $\omega_{L}$ represent the frequencies of
the upper and lower photonic Bloch modes.

While in an infinite Bragg reflector the Bloch wavevector $K$ has
a continuous spectrum, in our finite size system, $\omega_{U}$ and
$\omega_{L}$ are complex due to radiation losses and take discrete
values. These define the eigenmodes of the finite Bragg reflector
with frequencies  $\omega_{Ui}= \tilde{\omega}_{Ui}- i
\gamma_{Ui}$ and $\omega_{Li}= \tilde{\omega}_{Li}- i \gamma_{Li}$
on the upper and lower sides of the PBG, corresponding to minima
in the reflection spectrum outside the PBG \cite{beggs} (see
Figure 3). Here $\gamma_{Ui}$, $\gamma_{Li}$ denote respective
radiative decay rates which decrease with increasing number of
periods. Notably, modes with different $\omega_{Ui}$, are well
separated and orthogonal to each other and we thus consider
interaction of the exciton with only the first Bragg mode. The
exciton eigenfrequency is also complex: $\omega_{{\rm x}}=
\tilde{\omega}_{{\rm x}}- i \gamma_{{\rm x}}$ with $\gamma_{{\rm
x}}$ being the inverse of the exciton lifetime. For the conditions
of our experiments, when the exciton frequency $\omega_{{\rm x}}$
is in the vicinity of the upper photonic Bloch mode
$|B_{U}\rangle$, assuming $\Omega_{{\rm R}}<<\Omega_{{\rm B}}$,
the interaction between the exciton states $|X_{S}\rangle,
|X_{A}\rangle$, and the lower photonic state $|B_{L}\rangle$ can
be neglected. Recalling the definition of detuning $\Delta$ as the
difference between the upper edge of the PBG and the exciton
resonance frequency, $\Delta= \tilde{\omega}_{{\rm
U1}}-\tilde{\omega}_{{\rm x}}$, the Hamiltonian reads
\begin{equation}
\mathcal{H} =
 \left( \begin{array}{ccc}
\tilde{\omega}_{{\rm x}}+\Delta - i \gamma_{U1}& 0 &
\sqrt{2}\Omega_{{\rm R}}\\ 0 & \tilde{\omega}_{{\rm x}}- i
\gamma_{{\rm x}} & 0\\ \sqrt{2}\Omega_{{\rm R}}&0&
\tilde{\omega}_{{\rm x}}- i \gamma_{{\rm x}}
\end{array} \right) \label{hamiltonian4}
\end{equation}
and has three eigenfrequencies:
\begin{eqnarray}
\omega_{1} &=& \tilde{\omega}_{{\rm x}} - i \gamma_{{\rm x}}\\
\omega_{2,3} &=& \tilde{\omega}_{{\rm x}} + \frac{1}{2} \left (
\Delta - i \gamma_{+} \pm \sqrt{\Delta^{2}+ 8 \Omega_{{\rm R}}^{2}
- \gamma_{-}^{2} - 2 i \Delta \gamma_{-}} \right ) \nonumber \\
\end{eqnarray}
corresponding to the triplet of branches observed in the
experiments. Here, $ \gamma_{\pm} = \gamma_{U1} \pm  \gamma_{{\rm
x}}$. Therefore, the three modes result from the coupling of one
photonic mode to two excitonic ones. In Figure 2(c) the three
branches predicted from the coupled oscillator model are plotted
as a function of the detuning (grey solid lines), showing
excellent agreement with the experimental data (black circles).
\begin{figure}
\includegraphics[width=8.5cm]{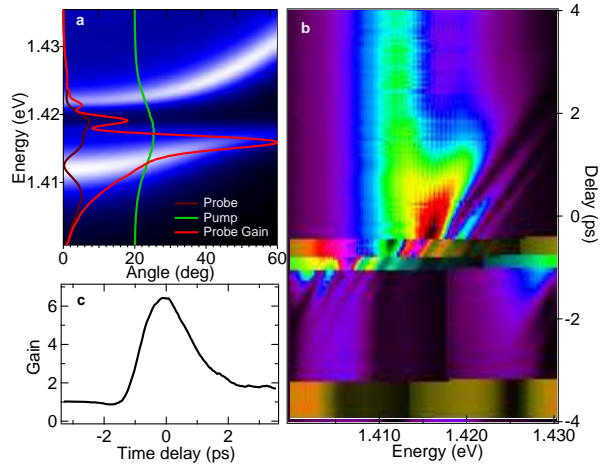}
\caption{(Color online) Bragg polariton amplification for pump
probe configuration shown in (a). Probe reflectivity spectra at
normal incidence (purple). 50mW pump pulse is resonant with the
lower polariton branch at 20$^{\circ}$ (green). Strong
amplification of the 230$\mu$W probe pulse is observed when the
pump beam is switched on (red line). (b) Contour plot on probe
reflectivity spectra at different time delays showing dispersive
gain peaks (c) Gain vs time delay showing ultrafast nature of the
amplification process.}
\end{figure}
Remarkably, the real part of the middle branch eigenfrequency
$\tilde{\omega}_{1}$ does not vary with detuning $\Delta$ since
the asymmetric photonic Bloch mode does not interact with the
symmetric exciton Bloch mode $|X_{S}\rangle$ \cite{panzarini}.
This can be directly envisioned when plotting in Figure 3(b) the
squared electric field at the QW sites calculated for the first
and second Bragg modes in the absence of the excitonic absorption.
As expected, the electric field envelope corresponds to a Bloch
wave exhibiting half a cycle within the structure. Furthermore, it
is seen that only a fraction of the structure's 30 QWs contribute
towards strong coupling, namely the ones positioned in the high
field region, whereas the rest in the low field region remain
uncoupled. This picture can differ substantially in structures
with narrow PBG where exciton coupling to both photonic Bloch
modes on either side of the PBG can occur. Such coupling mixes all
four modes and introduces dispersion to the middle branch
\cite{goldberg}. However, in the latter case the individual modes
become less pronounced as opposed to the sharp modes of this
study.

The investigated Bragg structure presents several important
advantages over the conventional strongly coupled microcavity
system: a) it ensures maximal overlap between the exciton
wavefunction and the light mode as opposed to microcavity in which
electric field penetration into the mirrors reduces Rabi splitting
b) constitutes a universal approach which allows incorporation of
wide variety of high refractive index contrast organic/inorganic
materials using layer by layer deposition c) offers an alternative
approach to achieving strong coupling regime in material systems
such as GaN where strain buildup poses serious limitations
\cite{baumberg}.

To confirm the potential of Bragg polaritons we explore their
nonlinear properties in angle-resonant polariton parametric
amplification geometry \cite{amplifier} exploiting their
microcavity-like dispersion relations [Fig. 4(a)]. We use
spectrally filtered 3.5meV bandwidth 1ps pump pulses to resonantly
excite lower polariton branch at the ``magic'' angle. Broadband
150fs probe pulses are incident normal to the sample and their
time-dependent reflectivity changes are recorded. In [Fig. 4(a)]
probe reflectivity spectra in the absence of the pump is
presented, clearly showing polariton dips at all three branches
(purple line). With the pump on the reflected probe spectrum at
zero time delay changes dramatically: a strong peak appears, blue
shifted by 3.5meV with respect to the bare lower polariton
reflectivity dip, and yields a peak gain of 6.4 [Fig. 4(c)]. While
this blue shift is larger than that generally observed in III-V
microcavities \cite{amplifier}, it remains smaller than half the
Rabi splitting. The temporal dynamics of the normalized probe
spectra [Fig.4(b)] show an ultrafast (1.8ps) response for the main
gain peak [Fig.4(c)], similar to parametric amplification in
conventional microcavities \cite{amplifier}. The well resolved
spectral fringes on the side of the main gain peak seen in
Fig.4(b) are ascribed to typical feature arising from the
energy-shifting resonance in time integrated spectra at fixed
pump-probe time delay (see \cite{Berstermann}). Finally, compared
to the case of strongly coupled planar microcavity, the gain peak
is broader (2.5meV) due to broader polariton resonances and
shorter polariton lifetimes.

In conclusion, we show that our unfolded microcavity structure
offers a novel laboratory for the exploration of the strong
coupling regime. Tuning of the characteristic mode of a Bragg
structure through the excitonic resonance produces strong coupling
with the appearance of mixed Bragg polariton states. Interpreting
the eigenmodes of the structure in terms of the normal modes of a
coupled harmonic oscillator system closely reproduces the Bragg
polariton branches. Our results suggest that Bragg polaritons
owing to their excitonic content can exhibit large nonlinearities.
Furthermore their polariton--like dispersion offers new
opportunities for the study of nonlinear phenomena such as
stimulated scattering, amplification and Bragg polariton lasing.

This research was supported by the Greek GSRT, project PENED 2003,
No: 03E$\Delta$816 and EU ITN projects ``CLERMONT 4" and
``ICARUS". The authors are thankful to A. V. Kavokin for
stimulating discussions.

\end{document}